\documentstyle[times,pramana,epsf,floats]{ias}
\begin{document}
\mark{{Half lives of spherical proton emitters.....}{P Roy Chowdhury, C Samanta and  D N Basu}}
\title{ Half lives of spherical proton emitters using density dependent M3Y interaction }

\author{P. Roy Chowdhury$^1$\thanks{E-mail:partha.roychowdhury@saha.ac.in}, C. Samanta$^{1,3}$ and D.N.Basu$^2$}

\address{Saha Institute of Nuclear Physics$^1$, Variable Energy Cyclotron Centre$^2$, 1/AF Bidhan Nagar, Kolkata 700064, India, Physics Dept., Virginia Commonwealth University$^3$, Richmond, VA23284-2000 USA}

\keywords{Proton radioactivity, Effective interaction, EOS, Incompressibility, Mass formula.}

\pacs{ 23.50.+z, 21.65.+f, 21.30.Fe, 25.55.Ci}

\abstract{The proton radioactivity lifetimes of spherical proton emitters from the ground and the isomeric states are calculated using the microscopic nucleon-nucleus interaction potentials. The daughter nuclei density distributions are folded with a realistic density dependent M3Y effective interaction supplemented by a zero-range pseudo-potential. The density dependence parameters of the interaction are extracted from the nuclear matter calculations. The saturation energy per nucleon used for nuclear matter calculations is determined from the co-efficient of the volume term of Bethe-Weizs\"acker mass formula which is evaluated by fitting the recent experimental and estimated atomic mass excesses from Audi-Wapstra-Thibault atomic mass table by minimizing the mean square deviation. The quantum mechanical tunneling probability is calculated within the WKB approximation. Spherical charge distributions are used for calculating the Coulomb interaction potentials. These calculations provide good estimates for the observed proton radioactivity lifetimes.}

\maketitle
\section{Introduction}
The discovery of proton radioactivity has provided a tool to obtain spectroscopic information \cite{Ba05}, \cite{Ab97} near the proton drip line. The decaying proton is the unpaired one not filling its orbit. These decay rates are sensitive to the proton decay $Q$ values and the orbital angular momenta which in turn help to determine the orbital angular momenta of the emitted protons. It has the lowest Coulomb potential among all possible charged clusters in nuclei and mass being the smallest it suffers the highest centrifugal barrier, enabling this process suitable to be dealt within WKB barrier penetration model. In the present work we provide estimates for the proton radioactivity lifetimes of some spherical proton emitters from the ground and the isomeric states using the nucleon-nucleus interaction potentials obtained microscopically by single folding the daughter nuclei density distributions with a realistic density dependent M3Y effective interaction (DDM3Y) whose density dependence parameters have been extracted from the nuclear matter calculations. 

\section{Energy coefficients of Bethe-Weizs\"acker mass formula from atomic mass excesses}

The Bethe-Weizsa\"cker formula \cite{Bw36} is an empirically refined form of the liquid drop model for the binding energy of nuclei. Expressed in terms of the mass number A and the atomic number Z for a nucleus, the binding energy B(A,Z) is given by,

\begin{equation}
 B(A,Z) = a_vA-a_sA^{2/3}-a_c\frac{Z(Z-1)}{A^{1/3}}-a_{sym}\frac{(A-2Z)^2}{A}+\delta,
\label{seqn1}
\end{equation}
\noindent
where $\delta=a_pA^{-1/2}$, for even N-even Z, $=-a_pA^{-1/2}$ for odd N-odd Z, $=0$ for odd A and N is the neutron number of the nucleus. Using the definition of nuclear binding energy B(A,Z) which is defined as the energy required to separate all the nucleons consituting a nucleus, the mass equation can be expressed  as 

\begin{equation}
 M_{nucleus}(A,Z) = Z m_p + (A-Z) m_n - B(A,Z),
\label{seqn2}
\end{equation}
\noindent
where $m_p$ and $m_n$ are the masses of proton and neutron respectively and $M_{nucleus}(A,Z)$ is the actual mass of the nucleus. The mass of the nucleus can be obtained from the atomic mass excess by use of the relationship

\begin{equation}
 M_{nucleus}(A,Z) = A u + \Delta M_{A,Z} - Zm_e  + a_{el} Z^{2.39} + b_{el} Z^{5.35},
\label{seqn3}
\end{equation}   
\noindent
where $m_e$ is the mass of an electron, the atomic mass unit u is 1/12 the mass of $^{12}C$ atom, $\Delta M_{A,Z}$ is the atomic mass excess of an atom of mass number A and atomic number Z and the electronic binding energy constants \cite{Lu03} $a_{el} = 1.44381 \times 10^{-5}$ MeV and $b_{el} = 1.55468 \times 10^{-12}$ MeV. Hence from Eq.(2) and Eq.(3) the atomic mass excess is given by 

\begin{equation}
 \Delta M_{A,Z} = Z \Delta m_H  + (A-Z) \Delta m_n - a_{el} Z^{2.39} - b_{el} Z^{5.35} - B(A,Z)
\label{seqn4}
\end{equation}
\noindent
where $\Delta m_H = m_p + m_e - u$ =  7.28897050 + $a_{el}$ + $b_{el}$ MeV and $\Delta m_n = m_n - u$ = 8.07131710 MeV.
Root mean square deviation $\sigma$ for atomic mass excesses is defined as 

\begin{equation}
 \sigma^2 =  (1/N) \sum [ \Delta M_{Th} - \Delta M_{Ex} ]^2
\label{seqn5}
\end{equation}   
\noindent
where the summation extends to $N$ data points for which experimental atomic mass excesses are known. The quantity $\Delta M_{Th}$ is theoretically calculated atomic mass excess obtained using Eq.(4) while $\Delta M_{Ex}$ is the corresponding experimental atomic mass excess obtained from the Audi-Wapstra-Thibault atomic mass table \cite{Au03}. For the calculations of the mean square deviation $\sigma^2$, masses \cite{Au03} of 3179 nuclei including the 951 extrapolated values which are predicted according to systematics, have been used for least square fitting. The values of the energy coefficients are $a_v=15.2602\pm0.0196 MeV$, $a_s=16.2673\pm0.0621 MeV$, $a_c=0.688974\pm0.001357 MeV$, $a_{sym}=22.2090\pm0.0481 MeV$, $a_p=10.0757\pm0.8536 MeV$. 

\section{The volume energy coefficient and the nuclear matter incompressibility }

A density dependent M3Y effective nucleon-nucleon (NN) interaction \cite{Sa79} based on the G-matrix elements of the Reid-Elliott NN potential has been used to determine the nuclear matter equation of state. The equilibrium density of the nuclear matter has been determined by minimizing the energy per nucleon. The density dependence parameters have been chosen to reproduce the saturation energy per nucleon and the saturation density of spin and isospin symmetric cold infinite nuclear matter, which is called the standard nuclear matter. The general expression for the DDM3Y interaction potential $v(s)$ is written as  

\begin{equation}
 v(s,\rho, \epsilon) = t^{M3Y}(s, \epsilon) g(\rho, \epsilon) = C t^{M3Y} (1 - \beta(\epsilon)\rho^{2/3}) 
\label{seqn6}
\end{equation}   
\noindent
where M3Y effective interaction potential supplemented by a zero range pseudopotential $t^{M3Y}$ is given by \cite{Ko84} 

\begin{equation}
  t^{\rm M3Y} = 7999 \frac{e^{ - 4s}}{4s} - 2134\frac{e^{- 2.5s}}{2.5s} + J_{00}(E) \delta(s)
\label{seqn7}
\end{equation}   
\noindent
where the zero-range pseudo-potential $J_{00}(\epsilon)$ representing the single-nucleon exchange term is given by 

\begin{equation}
 J_{00}(\epsilon) = -276 (1 - \alpha\epsilon) (MeV.fm^3)
\label{seqn8}
\end{equation}   
\noindent
The energy per nucleon $\epsilon$ obtained using the effective nucleon-nucleon interaction $v(s)$ for the standard nuclear matter, is given by

\begin{equation}
 \epsilon = [\frac{3\hbar^2k_F^2}{10m}] + \frac{g(\rho, \epsilon)\rho J_v}{2} = [\frac{3\hbar^2k_F^2}{10m}] + [\frac{\rho J_v C (1 - \beta\rho^{2/3})}{2}]
\label{seqn9}
\end{equation}   
\noindent
where m is the nucleonic mass, $k_F=(1.5\pi^2\rho)^{1/3}$ is the Fermi momentum, $\rho$ is the nucleonic density while $\rho_{0}$ being the saturation density for the standard nuclear matter and $J_v$ represents the volume integral of $t^{M3Y}$, the M3Y interaction supplemented by the zero-range pseudopotential. Eq.(9) can be differentiated with respect to $\rho$ to yield equation  

\begin{equation}
 \partial\epsilon/\partial\rho = [\hbar^2k_F^2/5m\rho] + J_v C [1 - (5/3)\beta\rho^{2/3}] /2 
\label{seqn10}
\end{equation}
\noindent
The equilibrium density of the nuclear matter is determined from the saturation condition $\partial\epsilon/\partial\rho = 0$. Then Eq.(9) and Eq.(10) with the saturation condition can be solved simultaneously for fixed values of the saturation energy per nucleon $\epsilon_0$ and the saturation density $\rho_{0}$ of the standard nuclear matter, to obtain the values of the density dependence parameters 

\begin{equation}
 \beta = [(1-p)\rho_{0}^{-2/3}]/[3-(5/3)p],~p = [10m\epsilon_0]/[\hbar^2k_{F_0}^2],
\label{seqn11}
\end{equation} 
\noindent

\begin{equation}
 C = -[2\hbar^2k_{F_0}^2] / [5mJ_v\rho_0(1 - (5/3)\beta\rho_0^{2/3})]~with~k_{F_0} = [1.5\pi^2\rho_0]^{1/3}, 
\label{seqn12}
\end{equation} 
\noindent
respectively. The incompressibility $K_0$ of the standard nuclear matter is given by  
  
\begin{equation}
 K_0 = k_F^2\partial^2\epsilon/\partial{k_F^2} = 9\rho^2\partial^2\epsilon/\partial\rho^2\mid_{\rho=\rho_0} = [-(\frac{3\hbar^2k_{F_0}^2}{5m}) - 5 J_v C \beta\rho_0^{5/3}]
\label{seqn13}
\end{equation}
\noindent 
Identifying $a_v$ as the saturation energy per nucleon for the spin and the isospin symmetric cold infinite nuclear matter and hence using the saturation energy per nucleon equal to $-15.2602 MeV$ along with the density dependent M3Y effective interaction and the commonly used values for saturation density equal to $0.1533 fm^{-3}$ \cite{Br90} and energy dependence parameter $\alpha=0.005 MeV^{-1}$, the density dependence parameters have been found to be $C=2.07$, $\beta=1.668 fm^2$ while the nuclear incompressibility is found to be $293.4 MeV$ which is in close agreement with experimental data \cite{Sc96} and theoretical estimates \cite{Br90}.

\section{The half lives of the spherical proton emitters}
The microscopic nuclear potentials $V_N(R)$ have been obtained by single folding the density of the daughter nucleus with the finite range realistic DDM3Y effective interacion as

\begin{equation}
 V_N(R) = \int \rho (\vec{r}) v[|\vec{r} - \vec{R}|] d^3r 
\label{seqn14}
\end{equation}
\noindent
where $\vec{R}$ and $\vec{r}$ are, respectively, the co-ordinates of the emitted proton and a nucleon belonging to the residual daughter nucleus with respect to its centre. The density distribution function $\rho$ used for the daughter nucleus, has been chosen to be spherically symmetric 

\begin{equation}
 \rho(r) = \rho_0 / [ 1 + exp( (r-c) / a ) ]
\label{seqn15}
\end{equation}                                                                                                                                       \noindent     
                        
\begin{equation}
where~~~~c = r_\rho ( 1 - \pi^2 a^2 / 3 r_\rho^2 ),~~r_\rho = 1.13 A_d^{1/3},~~a = 0.54~fm
\label{seqn16}
\end{equation}
\noindent
and the value of $\rho_0$ is fixed by equating the volume integral of the density distribution function to the mass number $A_d$ of the residual daughter nucleus. The distance s between any nucleon belonging to the residual daughter nucleus and the emitted proton is given by $s = |\vec{r} - \vec{R}|$ while the interaction potential between these two nucleons $v(s)$ appearing in Eq.(14) is given by the DDM3Y effective interaction. The total interaction energy $E(R)$ between the proton and the residual daughter nucleus is equal to the sum of the nuclear interaction energy, Coulomb interaction energy and the centrifugal barrier. Thus

\begin{equation}
 E(R) = V_N(R) + V_C(R) + \hbar^2 l(l+1) / (2\mu R^2)
\label{seqn17}
\end{equation}   
\noindent
where $\mu = m_p m_d/m_A$  is the reduced mass, $m_p$, $m_d$ and $m_A$ are the masses of the proton, the daughter nucleus and the parent nucleus respectively, all measured in the units of $MeV/c^2$. Assuming spherical charge distribution for the residual daughter nucleus, the proton-nucleus Coulomb interaction potential $V_C(R)$ is given by

\begin{equation}
 V_C(R) =(\frac{Z_d e^2}{2R_c}).[ 3 - (\frac{R}{R_c})^2]~for~R\leq R_c, = \frac{Z_d e^2}{R}~~otherwise
\label{seqn18}            
\end{equation}   
\noindent
where $Z_d$ is the atomic numbers of the daughter nucleus. The touching radial separation $R_c$ 
between the proton and the daughter nucleus is given by $R_c = c_p+c_d$ where $c_p$ and $c_d$ have been obtained using Eq.(16). The energetics allow spontaneous emission of protons only if the released energy $Q$ is a positive quantity. The half life of a parent nucleus decaying via proton emission is calculated using the WKB barrier penetration probability. The assault frequency $\nu$ is obtained from the zero point vibration energy $E_v = (1/2)h\nu$. The decay half life $T$ of the parent nucleus $(A, Z)$  into a proton and a daughter $(A_d, Z_d)$  is given by

\begin{equation}
 T = [(h \ln2) / (2 E_v)] [1 + \exp(K)]
\label{seqn19}
\end{equation}
\noindent
where the action integral $K$ within the WKB approximation is given by

\begin{equation}
 K = (2/\hbar) \int_{R_a}^{R_b} {[2\mu (E(R) - E_v - Q)]}^{1/2} dR
\label{seqn20}
\end{equation}
\noindent
where $R_a$ and $R_b$ are the two turning points of the WKB action integral determined from the equations 

\begin{equation}
 E(R_a)  = Q + E_v =  E(R_b).
\label{seqn21}
\end{equation}
\noindent
From a fit to the experimental data on cluster emitters a law given by eqn.(5) of ref.\cite{Po86}, which relates $E_v$ with $Q$, was found. For the present calculations the same law \cite{Po86} extended to protons for the zero point vibration energies along with the experimental $Q$ values \cite{Ba05}, \cite{Ab97} have been used. The same set of data of ref.\cite{Ab97} has been used for the present calculations and presented in table-1 below. In fig.1, the plots of theoretical and experimental proton radioactivity half lives have been presented. 

\begin{table}[h]
\caption{Comparison between experimentally measured and theoretically calculated half-lives of spherical proton emitters. The asterisk symbol (*) denotes the isomeric state. The experimental $Q$ values, half lives and $l$ values are taken from reference [1]. The results of the present calculations have been compared with the experimental values and with the results of UFM [1] and DWBA [2] estimates. }
\hskip4pc\vbox{\columnwidth=26pc}
\begin{tabular}{lllllllll}
Parent &Ang. &Released &Experimental &Present &UFM$^a$  \\ 
nucleus &mom.& Energy & log$_{10}$[half life] & calculations &DWBA$^b$& \\ \hline
$^A$Symb.& $l(\hbar)$ & $Q(MeV)$ &$log_{10}T(s)$ &$log_{10}T(s)$& $log_{10}T(s)$ \\ \hline
&&&&&&&&\\
$^{109}I$&2&0.829&-4.000&-4.25&-5.000$^b$\\ 
$^{112}Cs$&2&0.823&-3.301&-3.13&-4.167$^b$\\ 
$^{113}Cs$&2&0.977&-4.770&-5.53&-6.268$^b$\\ 
$^{146}Tm$&5&1.140&-0.629&0.52&-0.456$^b$\\ 
$^{146}Tm^*$&5&1.210&-1.143&-0.30&-1.276$^b$\\ 
$^{147}Tm$&5&1.071&0.591&0.99&1.095$^a$\\ 
$^{147}Tm^*$&2&1.139&-3.444&-3.38&-3.199$^a$\\ 
$^{150}Lu$&5&1.283&-1.180&-0.58&-0.859$^a$\\ 
$^{151}Lu$&5&1.255&-0.896&-0.65&-0.573$^a$\\ 
$^{156}Ta$&2&1.028&-0.620&-0.38&-0.461$^a$\\ 
$^{156}Ta^*$&5&1.130&0.949&1.67&1.446$^a$\\ 
$^{157}Ta$&0&0.947&-0.523&-0.43&-0.126$^a$\\ 
$^{160}Re$&2&1.284&-3.046&-2.99&-3.109$^a$\\ 
$^{161}Re$&0&1.214&-3.432&-3.46&-3.231$^a$\\ 
$^{161}Re^*$&5&1.338&-0.488&-0.58&-0.458$^a$\\ 
$^{165}Ir^*$&5&1.733&-3.469&-3.51&-3.428$^a$\\ 
$^{166}Ir$&2&1.168&-0.824&-1.10&-1.160$^a$\\ 
$^{166}Ir^*$&5&1.340&-0.076&0.23&0.021$^a$\\ 
$^{167}Ir$&0&1.086&-0.959&-1.27&-0.943$^a$\\ 
$^{167}Ir^*$&5&1.261&0.875&0.71&0.890$^a$\\ 
$^{171}Au^*$&5&1.718&-2.654&-3.03&-2.917$^a$\\ 
$^{185}Bi$&0&1.624&-4.229&-5.43&-5.184$^a$\\ 
$^{185}Bi^*$&5&1.611?&-4.357&-1.43&-1.678$^b$\\ 
\end{tabular} 
\end{table}

\begin{figure}[htbp]
\epsfxsize=8cm
\centerline{\epsfbox{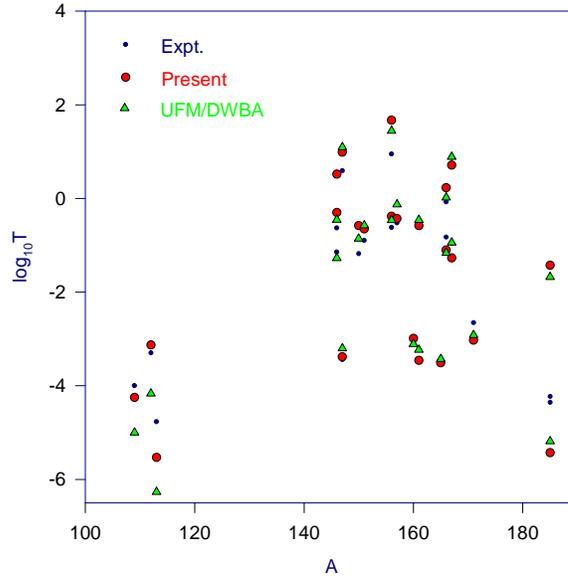}}
\caption{Comparision between theoretical and experimental proton radioactivity half lives. Dots represent the observed half lives while the circles and the triangles represent those corresponding to the present and the UFM[1]/DWBA[2] calculations respectively.}
\label{fig1}
\end{figure}

\section{Summary and conclusion}
The half lives for proton-radioactivity have been analyzed with microscopic nuclear potentials obtained by the single folding the DDM3Y effective interaction. The energy dependence parameters of the interaction have been obtained from the nuclear matter calculations. The saturation energy per nucleon used for nuclear matter calculations is determined by fitting the theoretical mass excesses based on semi-empirical liquid drop model to the experimental mass excesses. This procedure of obtaining nuclear interaction potentials has a profound theoretical basis. The results of the present calculations for the proton-radioactivity lifetimes are in good agreement over a wide range of experimental data.

\end{document}